\documentstyle[aps,twocolumn,epsfig]{revtex}
\begin{document}
\draft
\title{\large \bf RELATIVISTIC INSTANT--FORM APPROACH TO
THE STRUCTURE OF
TWO-BODY COMPOSITE SYSTEMS}

\author{A.F.Krutov\thanks{Electronic address:
krutov@ssu.samara.ru}}

\address {\it Samara  State University, 443011, Samara, Russia}

\author{V.E.Troitsky\thanks{Electronic address:
troitsky@theory.sinp.msu.ru}}
\address {\it D.V.Skobeltsyn Institute of Nuclear Physics,
Moscow State University, 119899, Moscow, Russia}

\date{December, 2001}
\maketitle

\begin{abstract}
A new approach to the electroweak properties of
two--particle composite
systems is developed. The approach is based on
the use of the instant form of relativistic Hamiltonian
dynamics.
The main novel feature of this
approach is the new method of construction of the matrix element
of the electroweak current operator.
The electroweak current matrix element satisfies the
relativistic covariance conditions and in the case of the electromagnetic
current also the conservation law automatically.
The properties of the system as well as the
approximations are formulated in terms of form factors.
The approach makes it possible to formulate relativistic
impulse approximation in such  a way that the Lorentz--covariance of
the current
is ensured. In the electromagnetic case the current conservation
law is also ensured.
Our approach gives good
results for the pion electromagnetic form factor in the whole
range of momentum transfers available for experiments at present
time, as well as for lepton decay constant of pion.

\end{abstract}
\narrowtext

\section{Introduction}
The
constructing of correct quantitative methods of calculation
for
composite--particle structure is an important line of
investigations in particle physics. In nonrelativistic dynamics
there exist different correct methods which use model or
phenomenological interaction potentials. However, in the case of
high energy one needs to develop relativistic methods. It is
worth noting that now the experiments on accelerators, in
particular at JLab are performed with such an accuracy that the
treatment of traditionally "nonrelativistic" systems (e.g. the
deuteron) requires to take into account relativistic effects.
Relativistic effects are important also in the treatment of
composite systems of light quarks.
However, the
relativistic treatment of hadron composite systems is a rather
complicated problem. Let us note that the use of the
methods of the field theory in this case encounters serious
difficulties. For example, it is well known that the
perturbative QCD can not be used in the case of quark bound
states (see, e.g.,~\cite{Gro93,Kei94W}).

In the present paper we will use the relativistic
constituent model which describes the hadron properties at quark
level in terms of degrees of freedom of constituent quarks. The
constituent quarks are considered as extended objects, the
internal characteristics of which (MSR, anomalous magnetic moments, form
factors) are parameters of model.
As a relativistic
variant of the constituent model we
choose the method of relativistic Hamiltonian dynamics (RHD)
(see, e.g.,
~\cite{LeS78,KeP91,Coe92,Kli98} and references therein).

The RHD method as a relativistic theory of composite systems
is based on the direct realization of the Poincar\'e algebra on
the set of dynamical observables on the Hilbert space.
RHD theory of particles lie
between local field theoretic models and nonrelativistic quantum
mechanical models.

Contrary to field theory, RHD is dealing
with finite number of degrees of freedom from the very
beginning. This is certainly a kind of a model approach. The
preserving of the Poincar\'e algebra ensures the relativistic
invariance.
So, the covariance of the description in the frame of RHD is due
to the existence of the unique unitary representation of
the inhomogeneous group $SL(2,C)$ on the Hilbert space of
composite system states with finite number of degrees of
freedom \cite{Nov75}.

The mathematics of RHD is similar to that of
nonrelativistic quantum mechanics and permits to assimilate
the sophisticated methods of phenomenological potentials and can
be generalized to describe three or more particles.

The idea of this approach --- RHD --- is originated by Dirac.
In ~\cite{Dir49} he considered different ways of description of
the evolution of classical relativistic systems --- different
forms of dynamics. Dirac  defined three main forms of
dynamics: point (PF),
instant (IF) and light--front (FF) dynamics.

RHD is based on the simultaneous action of two fundamental
principles: relativistic invariance and Hamiltonian principle
--- and presents the most adequate tool to treat the systems
with finite number of degrees of freedom.

Our aim is to construct a relativistic invariant approach to
electroweak structure of two--particle composite systems. The
main problem  here is the construction of the current operators
\cite{GrR87,ChC88,CoR94,Lev95,LeP98}.

It seems to us that RHD is the most adequate method for our
purpose. The use of RHD enables one to separate the main degrees
of freedom and thus to construct convenient models.

We use one of the forms of RHD, namely a version of the IF.

Our approach has a number of features that distinguish it from
other forms of dynamics and other approaches in the frames of
IF.

\begin{itemize}

\item
The electroweak current matrix element satisfies automatically the
relativistic covariance conditions and in the case of the electromagnetic
current also the conservation law.

\item
We propose a modified impulse
approximation (MIA). It is constructed in relativistically invariant
way. This means that our MIA does not depend on the choice of
the coordinate frame, and this contrasts principally with the
"frame--dependent" impulse approximation usually used in instant
form (IF) of dynamics.
\footnote{
It is known that correct impulse approximation (IA) realization
in the frame of traditional version of IF dynamics encounters
difficulties: the standard IA depends on the choice of
the coordinate frame. We show below that IA can be formulated in
an invariant way, the composite system form factors being
defined by the one--particle currents alone.}

\item
Our approach provides with correct and natural
nonrelativistic limit ("the correspondence principle" is
fulfilled).

\item
For composite systems (including the spin 1 case) the
approach guarantees the uniqueness of the solution for form factors
and it does not use such concepts as "good" and "bad" current components.

\end{itemize}

It is worth to notice that all known approaches (including the
perturbative quantum field theory (QFT)) encounter difficulties
while constructing a composite--system current operator
satisfying Lorentz--covariance and conservation conditions
\cite{GrR87,ChC88,CoR94,Lev95,LeP98}.

Similar difficulties arise in the frame of RHD approach
which is widely used in the theory of electroweak properties
of composite quark  and nucleon systems
~\cite{Kli98},~\cite{ChC88},~\cite{LeP98},
~\cite{GrK84,Jau91,KrT93,Sch94,CaG96,BaK96,YaO96,Kru97,AlK98,KrT99,BaK00,AlK00,And00,KrS00}.
At present time the FF dynamics
is the most developed and used for composite systems
\cite{ChC88,LeP98,GrK84,Jau91,Sch94,CaG96}.
However there are
some difficulties in the FF RHD approach
when the electroweak properties of composite systems are
considered. In particular, it was shown ~\cite{GrK84,Kei94} that
the calculated electromagnetic form factors for the systems with
the total angular momentum $J = 1$ (the deuteron, the $\rho$ --
meson) vary significantly with the rotation of the coordinate
frame. This ambiguity is caused by the breaking of the so called
angle condition ~\cite{GrK84,Kei94}, that is really by the
breaking of the rotation invariance of the theory. Some of the
difficulties of FF dynamics are discussed in ~\cite{Fud90}. A
possible way to solve the problem by adding some new
(nonphysical) form factors to the electromagnetic current was
proposed (see ~\cite{CaD98} and references therein).

A different approach to the problem was proposed recently in
Ref.~\cite{LeP98}, where a new method of construction of
electromagnetic current operators in the frame of FF dynamics
was given. The method of
~\cite{LeP98} gives unambiguous deuteron form factors. However,
as the authors of ~\cite{LeP98} note themselves, their
current operator and the one used in Ref.
~\cite{ChC88} are different, since both of them are obtained
from the free one, but in different reference frames, related by
an interaction dependent rotation.

Let us consider now the impulse approximation
(IA) which is widely used for the description of composite
systems. In IA a test particle interacts mainly with each
component separately, that is the electromagnetic current of the
composite system can be described in terms of one--particle
currents. In fact, the composite--system current is approximated
by the corresponding free--system current. This means that
exchange currents are neglected, or, in other words, that there
is no three--particle forces in the interaction of a test
particle with constituents. It is well known that the
traditional IA breaks the Lorentz--covariance of the
composite--system current and the conservation law for the
electromagnetic current (see, e.g., \cite{KeP91} for details).

To satisfy the conservation law in the frame of Bethe--Salpeter
equation and quasipotential equations, for example, it is
necessary to go beyond IA: one has to add the so called
two--particle currents to the current operator. In the case of nucleon
composite systems these
currents are interpreted as meson exchange currents
\cite{CoR94}.  In the case of deuteron this means the
simultaneous interaction of virtual
$\gamma$-- quanta with proton and neutron. However, in Ref.
\cite{VaD95} it is shown that the current conservation law can
be satisfied without such processes, although they contribute to
the deuteron form factor. It seems that at the present time
there is an intention to formulate IA with transformed
conservation properties without dynamical contribution of
exchange currents
\cite{LeP98,AlK00,CaD98}.

In the framework of the point form dynamics the current operator
was constructed in the Ref.\cite{Kli98}. The current operator in
\cite{Kli98} is Lorentz--covariant and the conservation law is
fulfilled. The approach is based on the realization of the
Wigner--Eckart theorem for the Poincar\'e group. The main idea
is to extract from the current matrix element the relativistic
invariant part -- the reduced matrix element, i.e. the form
factor, and to separate the covariant part. The form factors
contain all the dynamical information and the covariant part
describes the relativistic transformation properties of the
matrix element.

Our approach is a generalization of the method \cite{Kli98} for
the case of the instant form dynamics. However, the scenario of
the generalization of the Wigner-Eckart theorem is quite
different.

The IF of relativistic dynamics, although
not widely used, has some advantages.
The calculations can be performed in a natural straightforward
way without special coordinates. IF is particularly convenient
to discuss the nonrelativistic limit of relativistic results.
This approach is obviously rotational invariant, so IF is the
most suitable for spin problems.

We describe the dynamics of composite systems (the constituent
interaction) in the frame of general RHD axiomatics. However,
our approach differs from the traditional RHD by the way of
constructing of matrix elements of local operators. In
particular, our method of description of the electromagnetic
structure of composite systems permits the construction of
current matrix elements satisfying the Lorentz--covariance
condition and the current conservation law.

To construct the current operator in the frame of IF RHD we use
the general method of relativistic invariant parameterization of
matrix elements of local operators proposed as long ago as in
1963 by Cheshkov and Shirokov~\cite{ChS63}.

The method of ~\cite{ChS63} gives matrix elements of the
operators of arbitrary tensor dimension (Lorentz--scalar,
Lorentz-vector, Lorentz--tensor) in terms of a finite number of
relativistic invariant functions -- form factors. The form
factors contain all the dynamical information on the transitions
defined by the operator.

In the review \cite{KeP91}
two possible variants of such kind of representation of matrix
elements in terms of form factors are presented -- the
elementary--particle parameterization and the multipole
parameterization. The variant of parameterization given in
\cite{ChS63} is an alternative one. In \cite{ChS63} the authors
propose the construction of matrix elements in canonical basis
so it can be called canonical parameterization. This method was
developed for the case of composite systems in
~\cite{TrS69,KoT72}. The composite--system form factors in this
approach are in general case the distributions (generalized
functions), they are defined by continuous linear functionals on
a space of test functions.  Thus, for example, the current
matrix elements for composite systems are functionals, generated
by some Lorentz--covariant distributions, and the form factors
are functionals generated by regular Lorentz--invariant
generalized functions. We demonstrate these facts below, in
Sec.III, using a simple model as an example.

It is worth noting that the statement that the form factors
of a composite system are generalized functions is not something
exotic. This fact takes place also in the standard nonrelativistic
potential theory (see Sec.III(E)).

Our formalism also gives, in fact, the description of the
covariance properties of the operators in terms of
many--particle as well as one--particle currents. However, the
important feature of our formalism is the fact that form factors
or reduced matrix elements describing the dynamics of
transitions contain in the IA only the contributions of one--particle
currents.

So, our approach to the construction of the current operator
includes the following main points:

\noindent
1. We extract from the current matrix element of composite
system the reduced matrix elements (form factors) containing the
dynamical information on the process. In general these form
factors are generalized functions.

\noindent
2. Along with form factors we extract from the matrix element a
part which defines the symmetry properties of the current: the
transformation properties under Lorentz transformation, discrete
symmetries, conservation laws etc.

\noindent
3. The physical approximations which are used to calculate the
current are formulated not in terms of operators but in terms of
form factors.

In this paper we present the main points of
our approach. To make it transparent we consider here only
simple systems with zero total angular momenta, so that
technical details do not mask the essence of the method. We
demonstrate the effectiveness of the approach by calculating the
pion electroweak properties. In this case the canonical
parameterization is very simple and can be realized without
difficulties. The case of more complicated systems requires
rather sophisticated mathematics for canonical parameterization
of local operator matrix elements and will be considered
elsewhere.

The paper is organized as follows.
In Sect. II we remind briefly the basic statements of RHD,
especially of IF RHD.
The IF wave functions of
composite systems are defined.
 In Sect. III our approach to
relativistic theory of two--particle composite systems and their
electroweak properties is presented. A simple model is
considered in details: two spinless particles in the $S$-state
of relative motion, one of the particles being uncharged. The
electromagnetic form factor of the system is derived. The
standard conditions for the current operator are discussed. The
modified impulse approximation (MIA) is proposed. The results of
IA and MIA are compared. The nonrelativistic limit is considered.
In Sect.IV
the developed formalism is used in the case of the system of two
particles with spins 1/2. The pion electromagnetic form factor
and the lepton decay constant are derived. The model parameters
are discussed and the comparison of the results with the
experimental data is given. The results of calculations in IA
and MIA are compared and are shown to differ significantly. In
Sect.V the conclusion is given.

\section{Relativistic Hamiltonian Dynamics}

In this Section some basic equations of RHD
are briefly reviewed.

We use the so called instant form dynamics
(IF). In this form the kinematic subgroup
contains the
generators of the group of rotations and translations in
the three--dimensional Euclidean space (interaction independing generators):
\begin{equation}
 \hat{\vec J}\>,\quad\hat{\vec P}\;.
\label{kinem}
\end{equation}
The remaining generators are Hamiltonians (interaction
depending):
\begin{equation}
\hat P^0\>,\quad \hat{\vec N}\;.
\label{hamil}
\end{equation}

The additive including of interaction into the mass
square operator (Bakamjian--Thomas procedure \cite{BaT53},
see, e.g.,~\cite{KeP91} for details)
presents one of the possible technical ways to  include
interaction in the algebra of the Poincar\'e group:
\begin{equation}
\hat M_0^2 \to \hat M_I^2 = \hat M_0^2 + \hat U \;.
\label{M0toMI}
\end{equation}
Here $\hat M_0$ is the operator of invariant mass for
the free system and
$\hat M_I$ -- for the system with interaction. The interaction
operator $\hat U$ has to satisfy the following commutation relations:
\begin{equation}
\left [\hat {\vec P},\,\hat U\right ]
= \left[\hat {\vec J},\,\hat U\right ]
= \left [\vec\bigtriangledown_P,\,\hat U\right ] = 0\;.
\label{[PU]=0}
\end{equation}
These constraints (\ref{[PU]=0}) ensure that the
algebraic relations of Poincar\'e group
are fulfilled for interacting system.
The relations (\ref{[PU]=0}) mean that the interaction potential
does not depend on the total momentum of the system as well as
on the projection of the total angular momentum. This fact is
well established for a class of potentials, for example, for
separable potentials \cite{ItB90}. Nevertheless, the conditions
(\ref{M0toMI}) and (\ref{[PU]=0}) can be considered as the model
ones. There exists another approach
\cite{Shi59} where a potential depends on the total momentum
but that approach is out of scope of this paper.

In RHD the wave function of the system of interacting particles
is the eigenfunction of a complete set of commuting operators.
In IF this set is:
\begin{equation}
 {\hat M}_I^2\>,\quad
{\hat J}^2\>,\quad \hat J_3\>,\quad \hat {\vec P}\;.
\label{complete}
\end{equation}
${\hat J}^2$ is the operator of the square of the total
angular momentum. In IF the operators
${\hat J}^2\;,\;\hat J_3\;, \;\hat {\vec P}$
coincide with those for the free system. So, in
(\ref{complete}) only the operator
$\hat M_I^2$ depends on the interaction.

To find the eigenfunctions for the system
(\ref{complete}) one has first to construct the adequate basis
in the state space of composite system. In the case of
two-particle system (for example, quark-antiquark system
$q\,\bar q$) the Hilbert space in RHD is the direct product of
two one-particle Hilbert spaces:
${\cal H}_{q\bar q}\equiv {\cal H}_q\otimes {\cal H}_{\bar q}$.

As a basis in ${\cal H}_{q\bar q}$
one can choose the following set of two-particle state vectors:
$$
|\,\vec p_1\,,m_1;\,\vec p_2\,,m_2\,\!\rangle
= |\,\vec p_1\,m_1\,\!\rangle \otimes|\, \vec p_1\,m_2\,\rangle\;,
$$
\begin{equation}
\langle\,\!\vec p\,,m\,|\,\vec p\,'\,m'\,\!\rangle
= 2p_0\,\delta (\vec p - \vec p\,')\,\delta _{mm'}\;.
\label{p1m1p2m2}
\end{equation}
Here $\vec p_1 \;,\;\vec p_2$ are 3-momenta of particles,
$m_1\;,\;m_2$ --- spin projections on the axis $z$,
$p_0 = \sqrt {\vec p\,^2 +M^2}\;$,$\;M$ is the constituent mass.

One can choose another basis where the motion of the
two-particle center of mass is separated and where three
operators of the set
(\ref{complete}) are diagonal:
$$
|\,\vec P,\;\sqrt{s},\;J,\;l,\;S,\;m_J\,\rangle ,
$$
$$
\langle\,\vec P,\;\sqrt{s},\;J,\;l,\;S,\;m_J
|\,\vec P\,',\;\sqrt {s'},\;J',\;l',\;S',\;m_{J'}\,\rangle
$$
$$ = N_{CG}\,\delta^{(3)}(\vec P -\vec P\,')\delta(\sqrt{s} - \sqrt{s'})
\delta_{JJ'}\delta_{ll'}\delta_{SS'}\delta_{m_Jm_{J'}}\;,
$$
\begin{equation}
N_{CG} = \frac{(2P_0)^2}{8\,k\,\sqrt{s}}\>,\quad
k = \frac{1}{2}\sqrt{s - 4M^2}\;.
\label{PkJlSm}
\end{equation}
Here $P_\mu = (p_1 +p_2)_\mu$, $P^2_\mu = s$, $\sqrt {s}$
is the invariant mass of the two-particle system,
$l$ --- the orbital angular momentum in the center--of--mass frame (C.M.S.),
$\vec S\,^2=(\vec S_1 + \vec S_2)^2 = S(S+1)\;,\;S$
--- the total spin in C.M.S., $J$ --- the total angular
momentum with the projection $m_J$.

The basis (\ref{PkJlSm}) is connected with the basis
(\ref{p1m1p2m2}) through the Clebsh--Gordan (CG) decomposition
for the Poincar\'e group (see, e.g., \cite{KoT72}):
$$
|\,\vec P,\;\sqrt {s},\;J,\;l,\;S,\;m_J\,\rangle
$$
$$
= \sum _{m_1\>m_2}\,\int \,\frac {d\vec p_1}{2p_{10}}\,
\frac {d\vec p_2}{2p_{20}}\,|\,\vec p_1\,,m_1;\,\vec p_2\,,m_2\,\rangle
$$
\begin{equation}
\times
\langle\,\vec p_1\,,m_1;\,\vec p_2\,,m_2\,|
\,\vec P,\;\sqrt{s},\;J,\;l,\;S,\;m_J\,\rangle\;.
\label{Klebsh}
\end{equation}
Here
$$
\langle\,\vec p_1\,,m_1;\,\vec p_2\,,m_2\,|
\,\vec P,\;\sqrt {s},\;J,\;l,\;S,\;m_J\,\rangle
$$
$$
=\sqrt {2s}[\lambda (s,\,M^2,\,M^2)]^{-1/2}\,
2P_0\,\delta (P - p_1 - p_2)
$$
$$
\times\sum_{\tilde m_1\>\tilde m_2}
\langle\,m_1|\,D^{1/2}(p_1\,P)\,|\tilde m_1\,\rangle
\langle\,m_2|\,D^{1/2}(p_2\,P)\,|\tilde m_2\,\rangle
$$
$$
\times
\sum_{m_l\;m_S}\,\langle1/2\,1/2\,\tilde m_1\,\tilde m_2\,|S\,m_S\,\rangle
Y_{lm_l}(\vartheta \,,\varphi)
$$
$$
\times
\langle S\,l\,m_s\,m_l\,|Jm_J\rangle\;.
$$
Here $\lambda (a,b,c) = a^2 + b^2 + c^2 - 2(ab + bc + ac)$,
$Y_{lm_l}$ - a spherical harmonics,
$\vartheta \,,\varphi$ are the spherical angles of the
vector $\vec p = (\vec p_1 - \vec p_2)/2$ in the C.M.S.,
$\langle\,S\,m_S\,|1/2\,1/2\,\tilde m_1\,\tilde m_2\,\rangle$
and $\langle Jm_J|S\,l\,m_S\,m_l\,\rangle$ are the
CG coefficients for the group $SU(2)$,
$\langle\,\tilde m|\,D^{1/2}(P,p)\,|m\,\rangle$ -
the three--dimensional spin rotation matrix to be used
for correct relativistic invariant spin addition.

It is on the vectors (\ref{PkJlSm}), (\ref{Klebsh})
that the Poincar\'e--group representation is realized
in the vector state space of two free particles.
The vector in representation is determined by
the eigenvalues of the complete commuting set of operators:
\begin{equation}
\hat M^2_0 = \hat P^2\,,\;\hat J^2\,,\;\hat J_3\;.
\label{M0,J}
\end{equation}
The parameters $S$ and $l$
play the role of invariant parameters of degeneracy.

As in the basis
(\ref{PkJlSm}) the operators ${\hat J}^2\;,\;\hat J_3\;,\;\hat {\vec P}$
in (\ref{complete}) are diagonal, one needs to diagonalize only the operator
$\hat M_I^2$ in (\ref{complete}) in order to obtain the system wave function.

The eigenvalue problem for the operator ${\hat M}^2_I$ in the basis
(\ref{PkJlSm}) has the form of nonrelativistic Schr\"odinger
equation (see, e.g., \cite{KeP91}).

The corresponding composite--particle wave function has the form
$$
\langle\vec P\,',\,\sqrt {s'},\,J',\, l',\,S',\,m_J'|\,p_c\rangle  =
$$
\begin{equation}
=N_C\,\delta (\vec P\,' - \vec p_c)\delta _{JJ'}\delta _{m_Jm_J'}\,
\varphi^{J'}_{l'S'}(k')\;,
\label{wf}
\end{equation}
$$
N_C =
\sqrt{2p_{c0}}\sqrt{\frac{N_{CG}}{4\,k'}}\;.
$$
$|\,p_c\rangle$ is an eigenvector of the set (\ref{complete}); $J(J+1)$ and
$m_J$ are the eigenvalues of $\hat J^2\,,\;\hat J_3\,,$ respectively
(Eqs. (\ref{complete}), (\ref{M0,J})).

The two--particle wave function of relative motion
for equal masses and total angular momentum and total
spin fixed is:
\begin{equation}
\varphi^J_{lS}(k(s)) = \sqrt[4]{s}\,u_l(k)\,k\;,
\label{phi(s)}
\end{equation}
and the normalization condition has the form:
\begin{equation}
\sum_l\int\,u_l^2(k)\,k^2\,dk = 1\;.
\label{norm}
\end{equation}

Let us note that for composite quark systems one uses sometimes
instead of the equation (\ref{norm})
the following one:
\begin{equation}
 n_c\,\sum_l\int\,u_l^2(k)\,k^2\,dk = 1\;.
\label{norm nc}
\end{equation}
Here $n_c$ -- is the number of colours.
The wave function (\ref{phi(s)})
coincides with that obtained by "minimal relativization" in
\cite{FrG89}.
The normalization factors in
(\ref{phi(s)})
in this case correspond to the relativization obtained by the
transformation to relativistic density of states
\begin{equation}
k^2\,dk\quad\to\quad \frac{k^2\,dk}{2\sqrt{(k^2 + M^2)}}\;.
\label{rel den}
\end{equation}

The formalism of this Section is used in the next one to
present the method of calculation of electroweak properties of
composite systems. Particularly, the method of construction of
electroweak current operators is described.

\section{The new relativistic instant--form approach to the
electroweak structure of two body composite systems}

In this Section we present our approach to electroweak
properties of relativistic two--particle systems. To demonstrate
how one describes the electromagnetic properties of composite
systems in our version of the RHD instant form we first use the
following simple model. We consider the system of two
spinless particles in the
$S$-- state of relative motion, one particle having no charge.
Let us note that a similar model was used in
\cite{KeP91} where the authors gave the description of
constituent interaction in IF of RHD and obtained the mass
spectrum. The application of our method in general case
follows the scheme of this Section.
The case of $\pi$ - meson is investigated in Sec.IV and the $S=1$
case in \cite{KrT01}.

Electromagnetic properties of the system are determined by the
current operator matrix element. This matrix element is
connected with the charge form factor
$F_c(Q^2)$ as follows:
\begin{equation}
\langle p_c\,|j_\mu(0)|\,p'_c\,\rangle  = (p_c+p'_c)_\mu\,F_c(Q^2)\;,
\label{j=Fc}
\end{equation}
where $p'_c\;,\;p_c$ are 4--momenta of the composite system in
initial and final states,
$Q^2 = -t\;,\;q^2 = (p_c - p_c')^2 = t\;,\;q^2$ is the momentum--transfer
square.
The form ~(\ref{j=Fc}) is defined by the Lorentz covariance and
by the conservation law only and does not depend on the model
for the internal structure of the system.

The Eq.~(\ref{j=Fc}) presents the simplest example of the
extraction of a reduced matrix element, that is the simplest
realization of the Wigner--Eckart theorem on the Poincar\'e
group.  The 4--vector $(p_c + p'_c)_{\mu}$ describes symmetry
and transformation properties of the matrix element.  The
reduced matrix element (the form factor) contains all the
dynamical information on the process described by the current.
The representation of a matrix element in terms of
form factors often is referred to as the parameterization of
matrix element.  The scattering cross section for elastic
scattering of electrons by a composite system can be expressed
in terms of charge form factor $F_c(Q^2)$. So, form factor can
be obtained from experiment and it is interesting to calculate
it in a theoretical approach.

In this Section we calculate the form factor of our simple
composite system using the version of RHD IF based on the
approach of the Section II.

Now let us list the conditions for the operator of the conserved
electromagnetic current to be fulfilled in relativistic case
(see, e.g.,~\cite{Lev95}).\\
(i).{\it Lorentz--covariance}:
\begin{equation}
\hat U^{-1}(\Lambda )\hat j^\mu (x)\hat U(\Lambda ) =
\Lambda ^\mu_{\>\nu}\hat j^\nu (\Lambda ^{-1}x)\;.
\label{UjmuU}
\end{equation}
Here $\Lambda $ is the Lorentz--transformation matrix,
$\hat U(\Lambda ) $ -- the operator of the unitary
representation of the Lorentz group.\\
(ii).{\it Invariance under translation}:
\begin{equation}
\hat U^{-1}(a)\hat j^\mu(x)\hat U(a) = \hat j^\mu(x-a)\;.
\label{UajmuUa}
\end{equation}
Here $\hat U(a)$  is the operator of the unitary
representation of the translation group.\\
(iii).{\it Current conservation law}:
\begin{equation}
[\,\hat P_\nu\,\hat j^\nu(0)\,] = 0\;.
\label{Pj=0}
\end{equation}
In terms of matrix elements
$\langle\,\hat j^\mu(0)\,\rangle$
the conservation law can be written in the form:
\begin{equation}
q_\mu\,\langle\,\hat j^\mu(0)\,\rangle  = 0\;.
\label{Qj=0}
\end{equation}
Here $q_\mu$ is 4-vector of the momentum transfer.\\
(iv).{\it Current--operator transformations under space--time
reflections are}:
$$
\hat U_P\left(\,\hat j^0(x^0\,,\vec x)\,,\hat{\vec j}(x^0\,,\vec x)\right)
\hat U^{-1}_P
$$
$$
=
\left(\,\hat j^0(x^0\,,-\,\vec x)\,,-\,\hat{\vec j}(x^0\,,-\,\vec x)\right)\;,
$$
\begin{equation}
\hat U_R\,\hat j^\mu(x)\,\hat U^{-1}_R = \hat j^\mu(-\,x)\;.
\label{UpUr}
\end{equation}
In (\ref{UpUr}) $\hat U_P$ is the unitary operator for the
representation of space reflections and
$\hat U_R$ is the antiunitary operator of the representation of
space-time reflections
$R = P\,T$.\\
(v).{\it Cluster separability condition}:
If the interaction is switched off then the current operator
becomes equal to the sum of the operators of one--particle
currents.\\
(vi).{\it The charge is not renormalized by the interaction
including}:  The electric charge of the system with interaction
is equal to the sum of the constituent electric charges.

In this paper the explicit equations for the form factors are
obtained taking into account all the listed conditions.

\subsection* {A. Electromagnetic properties of the system of
free particles}

Let us consider first the simple two--particle system described in the
beginning of Section III.
Electromagnetic current $j^{(0)}_\mu(0)$ of the two--particle
free system
can be calculated in the
representation given by the basis
(\ref{p1m1p2m2}) or in the representation given by the basis
(\ref{PkJlSm}).
In the first case the operator has the form
$j^{(0)}_\mu = j_{1\mu}\otimes I_{2}$. Here
$j_{1\mu}$ is the electromagnetic current of the charged particle and
$I_{2}$ is the unity operator in the Hilbert space of
states of the uncharged particle.
$$
\langle\vec p_1;\vec p_2|j_\mu^{(0)}(0)|\vec p\,'_1;\vec p\,'_2\rangle
$$
\begin{equation}
= \langle\vec p_2|\vec p\,'_2\rangle
\langle\vec p_1|j_{1\mu}(0)|\vec p\,'_1\rangle\;.
\label{j=j1xI}
\end{equation}
The matrix element of the one spinless particle current
in the free case contains only one form factor -- the charge form
factor of the charged particle
$f_1(Q^2)$:
\begin{equation}
\langle\vec p_1|j_{1\mu}(0)|\vec p\,'_1\rangle  =
(p_1+p'_1)_\mu\,f_1(Q^2)\;.
\label{j=f1}
\end{equation}

So, the electromagnetic properties of the system of two free
particles
(\ref{j=Fc}) are defined by the form factor
$f_1(Q^2)$, containing all the dynamical information on elastic
processes described by the matrix element
(\ref{j=j1xI}) \cite{KeP91}.
Particularly, the charge of the system is defined by the value
of this form factor at $Q^2\to 0$:
\begin{equation}
\lim_{Q^2\to 0}\,f_1(Q^2) = f_1(0) = e_c\;.
\label{f1(0)}
\end{equation}
$e_c$ is the system charge.

Now let us write the electromagnetic--current matrix element
for the two--particle free system in the basis where the
center--of--mass motion is separated (\ref{PkJlSm}):
\begin{equation}
\langle\vec P,\sqrt s,\mid j_\mu^{(0)}(0) \mid \vec P',\sqrt{s'}\rangle\;.
\label{*}
\end{equation}
Here the variables which take zero values are omitted:
$J = S = l = 0$. One can consider the matrix element
(\ref{*}) as a matrix element of an irreducible tensor operator
on the Poincar\'e group and one can use the Wigner--Eckart
theorem, i.e. the canonical parametrization ~\cite{ChS63},
~\cite{TrS69},~\cite{KoT72}
giving a technical realization of this theorem.
Thus, one can write
(\ref{*}) in the form
$$
\langle\vec P,\sqrt s,\mid j_\mu^{(0)}(0) \mid \vec P',\sqrt{s'}\rangle
$$
$$
= A_\mu (s,Q^2,s')\;\langle \sqrt{s}||g_0(Q^2)||\sqrt{s'}\rangle
$$
\begin{equation}
= A_\mu (s,Q^2,s')\;g_0 (s,Q^2,s')\;.
\label{j=A mu g0}
\end{equation}
It is easy to understand
the motivation for the parameterization
(\ref{j=A mu g0})
for our simple
system. The 4--vector $A_{\mu}$ describes the transformation
properties of the matrix element and the invariant function
$g_0 (s,Q^2,s')$ contains the dynamical information on the
process. We will refer to $g_0 (s,Q^2,s')$ as to free
two--particle form factor. For more complicated systems the
parameterization corresponding to the Wigner--Eckart theorem
for the Poincar\'e group can be performed using a special
mathematical techniques as described in the papers
~\cite{ChS63},~\cite{KoT72}, ~\cite{KrT01}.

So $A_{\mu} (s,Q^2,s')$ is defined by the
current transformation properties (the Lorentz--covariance and
the conservation law):
\begin{equation}
A_\mu =\frac{1}{Q^2}[(s-s'+Q^2)P_\mu + (s'-s+Q^2) P\,'_\mu]\;.
\label{Amu}
\end{equation}

Thus, in the basis (\ref{PkJlSm}) the electromagnetic properties
of the free two--particle system are defined by the free two--
particle form factor
$g_0 (s,Q^2,s')$.

So, in both representations (defined by the basis
(\ref{p1m1p2m2}) as well as by the basis (\ref{PkJlSm}))
we pass from the description of the system in terms of matrix
elements to that in terms of Lorentz--invariant form factors.

One can see that (\ref{j=j1xI}) and
(\ref{j=A mu g0}) describe electromagnetic properties in terms
of only one form factor. Both of these descriptions are,
certainly, equivalent from the physical point of view.
Let us consider the difference between these descriptions.
As we will show below by direct calculation the free
two--particle form factor
$g_0 (s,Q^2,s')$ is not an ordinary function but has to be
considered in the sense of distributions
in variables $s\;,\;s'$, generated by a locally integrable
function. So, $g_0 (s,Q^2,s')$ is a regular
generalizad function.
All the properties of
$g_0(s,Q^2,s')$  have to be considered as the properties of a
functional given by the integral over the variables
$s\;,\;s'$ of the function $g_0(s,Q^2,s')$
multiplied by a test function. As test functions it is
sufficient to take a large class of smooth functions that give
the uniconvergence of the integral. In particular, the limit
(\ref{f1(0)}) giving the total charge of the system through
two--particle form factor is now the weak limit:
\begin{equation}
\lim_{Q^2\to 0}\langle g_0(s,Q^2,s')\,,\phi(s,s')\rangle .
\label{g0(0)}
\end{equation}
Here $\phi(s,s')$ is a function from the space of test
functions. The precise definition of the functional will
be given below.

At the first glance it seems that the description of the
two--particle free system in terms of the form factor
$g_0(s,Q^2,s')$ is too complicated. However, so is the reality,
as we will see later in the Subsection III(E). In fact, this kind
of description is used implicitly for a long time in
nonrelativistic theory of composite systems,
without calling things by their proper names. It is this kind
of description that makes it possible to construct the
electromagnetic current operator with correct transformation
properties for interacting systems.

The locally integrable function
$g_0(s,Q^2,s')$ can be easily obtained by use of CG
decomposition (\ref{Klebsh}) for the Poincar\'e group.
Using (\ref{Klebsh})  we obtain for (\ref{j=A mu g0}):
$$
\langle\vec P,\sqrt{s},\mid j_\mu^{(0)}(0) \mid \vec P',\sqrt{s'}\rangle
$$
$$
=  \int\,\frac{d\vec p_1}{2\,p_{10}}
\frac{d\vec p_2}{2\,p_{20}}\frac{d\vec p_1\,'}{2\,p'_{10}}
\frac{d\vec p_2\,'}{2\,p'_{20}}\,
\langle\vec P,\sqrt s,\mid \vec p_1;\vec p_2\rangle
$$
\begin{equation}
\times\langle\vec p_1;\vec p_2|j_\mu^{(0)}(0)|\vec p\,'_1;\vec p\,'_2\rangle
\langle\,\vec p\,'_1;\vec p\,'_2|\vec P',\sqrt{s'}\rangle\;.
\label{jP=jp1p2}
\end{equation}
To calculate the free two--particle form factor one has to use
(\ref{j=j1xI}), (\ref{j=f1}), (\ref{j=A mu g0})
and the explicit form of CG coefficients
(\ref{Klebsh}) for quantum numbers of the system.
As the particles of the system under consideration are spinless,
now (\ref{Klebsh}) does not contain $D$ -- functions.

It is convenient to integrate in (\ref{jP=jp1p2})
using the coordinate frame with
$\vec P\,'=0\;,\;\vec P =(0,0,P)$.
As the result we obtain the following relativistic invariant
form for the function
$g_0(s,Q^2,s')$:
$$ g_0(s,Q^2,s')
$$
\begin{equation}
= \frac{(s+s'+Q^2)^2\,Q^2}{2\,\sqrt{(s-4M^2) (s'-4M^2)}}\>
\frac{\vartheta(s,Q^2,s')}{{[\lambda(s,-Q^2,s')]}^{3/2}}\,f_1(Q^2)\;.
\label{ff-nonint}
\end{equation}
Here $\vartheta(s,Q^2,s')=
\theta(s'-s_1)-\theta(s'-s_2)$, and $\theta$ is the step
function.
The result, naturally, does not depend on the choice of the
coordinate frame.

$$
s_{1,2}=2M^2+\frac{1}{2M^2} (2M^2+Q^2)(s-2M^2)
$$
$$
\mp \frac{1}{2M^2}\sqrt{Q^2(Q^2+4M^2)s(s-4M^2)}\;.
$$
The functions $s_{1,2}(s,Q^2)$ give
the kinematically available region
in the plane
$(s,s')$ (see ~\cite{TrS69}).

One can see that the free two--particle form factor
$g_0(s,Q^2,s')$ (\ref{ff-nonint}) has in fact to be interpreted
in terms of the distributions: The ordinary limit as
$Q^2\to$ 0 is zero because of the cutting
$\vartheta$ -- functions and the static limit exists only as
the weak limit (\ref{g0(0)}).

Let us calculate this limit. Let us define the functional giving
regular generalized function as a functional in
{\bf R}$^2$ as follows:
$$
\langle g_0 (s,Q^2,s')\,,\phi(s,s')\rangle
$$
\begin{equation}
= \int\,d\mu(s,s')\,g_0 (s,Q^2,s')\,\phi(s,s')\;.
\label{<>}
\end{equation}
Here
$$
d\mu(s,s') =
16\,\sqrt[4]{ss'}\,
\theta(s - 4\,M^2)\,\theta(s' - 4\,M^2)\,d\mu(s)\,d\mu(s')\;,
$$
\begin{equation}
d\mu(s) = \frac{1}{4}k\,d\sqrt{s}\;.
\label{dmu}
\end{equation}
The $\theta$ -- functions in these formula
give the physical region of possible variations
of the invariant mass squares in the initial and final states
explicitly.
The measure
(\ref{dmu}) is due to the relativistic density of states
(\ref{phi(s)}), (\ref{rel den}).  $\phi(s\,,s')$
is a function from the test function space. So, for example, the
limit of $g_0(s\,,Q^2\,,s')$ as $Q^2\;\to\;$0  (the static
limit)  has the meaning only as the weak limit
(compare with (\ref{f1(0)})):
\begin{equation}
\lim_{Q^2\to 0}\langle\,g_0,\phi\,\rangle
= \langle{e}_á \delta(\mu(s') - \mu(s)), \phi\,\rangle\;.
\label{Q2=0}
\end{equation}
It is this weak limit that gives the electric charge of the free
two--particle system. If the test functions are normalized with
the relativistic density of states, then the r.h.s. of the Eq.
(\ref{Q2=0}) is equal to the total charge of the system.


\subsection*{B.
Electromagnetic structure of the system of two interacting
particles.}

Now let us consider the electromagnetic structure of our simple
model (\ref{j=Fc}) in the case of interacting particles.

As we have mentioned in Sec.II when constructing the bases
(\ref{p1m1p2m2}) and (\ref{PkJlSm}) in the frame of RHD the
state vector
$|\,p_c\,\rangle $ belongs to the direct product of two
one--particle spaces. We can write the decomposition of this
vector with $J=l=S=m_J=$0 in the basis
(\ref{PkJlSm}). Now (\ref{j=Fc}) has the form:
$$
\int\,\frac{d\vec P\,d\vec
P\,'}{N_{CG}\,N_{CG}'}\,d\sqrt{s}\,d\sqrt{s'}\,
\langle p_c|\vec P\,,\sqrt{s}\,\rangle
\langle\vec P\,,\sqrt{s}|j_\mu(0)|\vec P\,'\,,\sqrt{s'}\rangle
$$
\begin{equation}
\times
\langle\vec P\,'\,,\sqrt{s'}|p_c'\rangle  = (p_c+p'_c)_\mu\,F_c(Q^2)\;.
\label{int=Fá}
\end{equation}
Here
$\langle\vec P\,'\,,\sqrt{s'}|p_c'\rangle$
is the wave function in the sense of the instant form of RHD
(\ref{wf}).

Using (\ref{wf}) we obtain for (\ref{int=Fá}):
$$
\int\,\frac{N_c\,N'_c}{N_{CG}\,N_{CG}'}\,d\sqrt{s}\,d\sqrt{s'}\,
\varphi(s)\,\varphi(s')
$$
\begin{equation}
\times \langle\vec p_c\,,\sqrt{s}|j_\mu(0)|\vec p_c\,'\,,\sqrt{s'}\rangle
= (p_c+p'_c)_\mu\,F_c(Q^2)\;.
\label{int ds=Fc}
\end{equation}
We have omitted in the wave function
(\ref{phi(s)}) the variables with zero values:
$J=S=l=$0.

Let us discuss the possibility of using the
Wigner--Eckart theorem (or the canonical parametrization)
in the case of the matrix element $\langle\vec
p_c\,,\sqrt{s}|j_\mu(0)|\vec p_c\,'\,,\sqrt{s'}\rangle$ in
(\ref{int ds=Fc}).  In the previous cases the state vectors and
the operators entering matrix elements transformed following one
and the same representation of the nonuniform group $SL(2,C)$
\cite{Nov75}. Let us perform the Lorentz transformation of the
current operator:
\begin{equation} \hat U^{-1}(\Lambda)j^\mu(0)\hat U(\Lambda) =
\tilde j^\mu(0)\;.  \label{UjU=tj} \end{equation}
We obtain:
$$ \langle p_á|\tilde
j^\mu(0)|p_á'\rangle = \langle p_á|\hat
U^{-1}(\Lambda)j^\mu(0)\hat U(\Lambda)|p_á'\rangle $$
\begin{equation}
= \langle \Lambda p_á|j^\mu(0)|\Lambda p_á'\rangle\;.
\label{tj=LpjLp}
\end{equation}
This means that the transformation properties of the current
4--vector
(\ref{UjmuU}) can be described using 4--momenta of the initial
and final states, i.e. one can use the canonical
parametrization.

In the matrix element in the integrand of
(\ref{int ds=Fc}) the state vectors and the operator transform
following the different representations of the group $SL(2,C)$. The
current operator describes the transitions in the system of two
interacting particles and transforms following the
representation with the generators of Lorentz boosts depending
on the interaction (\ref{complete}). The state vectors belong to
the basis (\ref{PkJlSm}) and physically describe the system of
two free particles and, so, transform following a representation
with generators which do not depend on the interaction
(\ref{M0,J}).
So, if one considers the matrix element
$\langle\vec p_c\,,\sqrt{s}|j_\mu(0)|\vec p_c\,'\,,\sqrt{s'}\rangle$
(that is the interaction current between free states) {\it per
se} , not in the context of the decomposition
(\ref{int ds=Fc}), one can not use the Wigner--Eckart theorem.

However, one must consider this matrix element as a generalized
function, that is as having meaning only as the integrand in
(\ref{int ds=Fc})). Let us show that in this case one can use
the Wigner--Eckart theorem.

The set of the free two-particle states
$|\vec P\,,\sqrt{s}\rangle\;$ is complete:
\begin{equation}
\hat I = \int\,\frac{d\vec P}{N_{CG}}\, d\sqrt{s}\,
|\vec P\,,\sqrt{s}\rangle\langle\vec P\,,\sqrt{s}|\;.
\label{I=compl}
\end{equation}
Using (\ref{wf}), (\ref{I=compl}) we obtain :
$$ \int\,\frac{N_c\,N_c'}{N_{CG}\,N_{CG}'}\,
d\sqrt{s}\,d\sqrt{s'}\,
\varphi(k)\varphi(k')
$$
$$
\times
\langle\vec p_c\,,\sqrt{s}|\hat U^{-1}(\Lambda)j_\mu(0)\hat U(\Lambda)
|\vec p_c\,'\,,\sqrt{s'}\rangle
$$
$$
= \langle\,p_c|\hat U^{-1}(\Lambda)\,\hat I\,j_\mu(0)\,\hat I\,\hat U(\Lambda)
|\,p_c\,'\rangle
$$
$$
= \langle\Lambda p_c\,|\hat I\,j_\mu(0)\,\hat I|\Lambda p_c\,'\rangle
$$
$$
= \int\,\frac{N_c\,N_c'}{N_{CG}\,N_{CG}'}\,
d\sqrt{s}\,d\sqrt{s'}\,\varphi(k)\varphi(k')
$$
\begin{equation}
\times\langle\Lambda\vec p_c\,,\sqrt{s}|j_\mu(0)|
\Lambda\vec p_c\,'\,,\sqrt{s'}\rangle\;.
\label{j=int3}
\end{equation}

So, we have an analog of
(\ref{tj=LpjLp}) {\it in the sense of distributions}
and we can use the Wigner--Eckart theorem in the intergrand.
One can speak about the Wigner--Eckart theorem in weak sense.

Now the problem of canonical parametrization of the matrix
element (\ref{int ds=Fc}) can be solved if one consider the
equality (\ref{int ds=Fc}) as the equality of two functionals.

Using (\ref{phi(s)}), (\ref{dmu}) we can rewrite (\ref{int ds=Fc})
in the form of the functional in {\bf R}$^2$:
$$
\int\,d\mu(s\,,\,s')\,
u(k(s))\,J_\mu(\vec p_c\,,\sqrt{s};\vec p_c\,'\,,\sqrt{s'})\,u(k(s'))
$$
\begin{equation}
= (p_c+p'_c)_\mu\,F_c(Q^2)\;,
\label{int dmu=Fá}
\end{equation}
$$
J_\mu(\vec p_c\,,\sqrt{s};\vec p_c\,'\,,\sqrt{s'}) =
\frac{N_c\,N'_c}{N_{CG}\,N_{CG}'}
\langle\vec p_c\,,\sqrt{s}|j_\mu|\vec p_c\,'\,,\sqrt{s'}\rangle\;.
$$

The l.h.s. in (\ref{int dmu=Fá}) contains a
functional in
{\bf R}$^2$ generated by the Lorentz--covariant function (current
matrix element). Let us denote
\begin{equation}
\psi(s\,,\,s') = u(k(s))\,u(k'(s'))\;.
\label{psi(ss')}
\end{equation}
The functional in the l.h.s. of (\ref{int dmu=Fá})
is given on the set of test functions
$\psi(s\,,\,s')$ through an integral in {\bf R}$^2$
and defines a Lorentz--covariant (regular) generalized function
with the values in the Minkowski space
(see, e.g., \cite{BoL87}). Here $Q^2$ is a parameter.
The test--function space can be (in general) larger than
(\ref{psi(ss')}). However, the uniconvergence of
(\ref{int dmu=Fá}) has to be guaranteed.

Let us write the matrix element in the form analogous to
(\ref{j=A mu g0}):
\begin{equation}
J_\mu(\vec p_c\,,\sqrt{s};\vec p_c\,'\,,\sqrt{s'})
= B_\mu(s\,,Q^2\,,s')G(s\,,Q^2\,,s')\;.
\label{j=BG}
\end{equation}
The covariant part in (\ref{j=BG}) (as well as in
(\ref{j=A mu g0})), the vector $B_\mu(s\,,Q^2\,,s')$,
is supposed to be an ordinary smooth function and the invariant
part
$G(s\,,Q^2\,,s')$ is generalized function. In fact,
$G(s\,,Q^2\,,s')$ is the reduced matrix element containing the
information on the process. This kind of representation of a
Lorentz covariant generalized function as a product of a Lorentz
covariant ordinary smooth function and a Lorentz invariant
generalized function was described in
\cite{BoL87}.

Using (\ref{j=BG}) we can rewrite (\ref{int dmu=Fá}) in the
following form:
$$
\int\,d\mu(s\,,\,s')\,\psi(s\,,s')B_\mu(s\,,Q^2\,,s')G(s\,,Q^2\,,s')
$$
\begin{equation}
= (p_c + p'_c)_\mu\,F_c[\psi](Q^2)\;.
\label{int BG=Fá}
\end{equation}
To obtain the vector $B_\mu$ let us require the
Eq.(\ref{int BG=Fá}) to be covariant in the sense of
distributions, that is to be valid for any test function
$\psi(s\,,s')$ in any fixed frame.
The variation of test function in the functional
(\ref{int BG=Fá}) means in fact, following
(\ref{psi(ss')}), the variation of the wave function of the
internal motion. Under such a variation the vector in the r.h.s
of (\ref{int BG=Fá}) is unchanged as it is constructed with
4--vectors describing the motion of the system as a whole,
independent of the internal constituent motion. As to the form
factor in the r.h.s it varies under the test function variation.
So, under a variation of the test function the r.h.s. of
(\ref{int BG=Fá}) remains to be collinear to the vector
$(p_c + p'_c)_\mu$. At the same time, under arbitrary variation
of the test function the vector in the l.h.s. in general changes
the direction. So, for the validity of the equality
(\ref{int BG=Fá}) with arbitrary test function it is sufficient
to require that the following equation
\begin{equation}
B_\mu(s\,,Q^2\,,s') = (p_c + p'_c)_\mu
\label{B}
\end{equation}
holds. This choice of the vector $B_\mu$ in (\ref{B})
ensures that the l.h.s. of
(\ref{int dmu=Fá}) satisfies the condition of Lorentz covariance
for the current as well as the condition of current
conservation.

Let us discuss the physical meaning of the representation
(\ref{j=BG}), (\ref{B}) for the matrix element. As this
representation is explicitly Lorentz covariant and also
satisfies the current conservation law, then it means that the
current operator for the composite system contains the
contribution not only of one--particle currents but of
two--particle currents, too
(see, e.g., \cite{KeP91}):
\begin{equation}
j = \sum_{k}\,j^{(k)} + \sum_{k<m}\,j^{(km)}\;.
\label{j=jk+jkm}
\end{equation}
Here the first term is the sum of one--particle currents and the
second -- of two--particle currents. In the case of our simple
model each sum in (\ref{j=jk+jkm})
contains only one term. It is well known that if one approximates
$j(x) \approx \sum_{k}\,j^{(k)}(x)$  then the current operator
in IF dynamics does not satisfy the condition of Lorentz
covariance and the conservation law
~\cite{KeP91}. So, from the physical point of view, the
covariant part of the current matrix element
(\ref{B}) which defines the transformation properties of the
current in (\ref{int dmu=Fá}) is given by
(\ref{j=jk+jkm}) and contains the contributions of one-- and
two-- particle currents.

The invariant part of the decomposition
(\ref{j=BG}) is the form factor or the reduced matrix element
$G(s\,,Q^2\,,s')$ and contains the information on the dynamics of
the scattering of test particle by each of the constituents
(the first term in (\ref{j=jk+jkm})), i.e. by the free
two--particle system, as well as by two constituent
simultaneously (the second term). So, the form factor contains
the contribution of the free system form factor
(\ref{ff-nonint}) and the contribution of some exchange currents
analogous to meson currents in nucleon systems \cite{GrR87}.

\begin{equation}
G(s\,,Q^2\,,s') = g_0(s,Q^2,s') + G_c(s\,,Q^2\,,s')\;.
\label{G=g0+Gc}
\end{equation}
Here $G_c$ is the reduced matrix element containing the
contribution of two--particle currents
~(\ref{j=jk+jkm}).

Using
(\ref{phi(s)}), (\ref{dmu}), (\ref{psi(ss')}), (\ref{B})
one can obtain from (\ref{int BG=Fá}) the scalar equation
of the following form:
\begin{equation}
\int\,d\sqrt{s}\,d\sqrt{s'}\,
\varphi(s)\,G(s\,,Q^2\,,s')\, \varphi(s') = F_c(Q^2)\;.
\label{int G=Fá}
\end{equation}
The representation
(\ref{int G=Fá}) for the charge form factor of the system is
quite general.

Let us note that one can use the described formalism in general
case of composite systems with nonzero total angular momentum
$J$ (the detailed consideration is given in
\cite{KrT01}).  In this case the current matrix element in the
(\ref{int ds=Fc})--like decomposition is a matrix with matrix
indices being the total angular momentum projections
$m_J'\,,m_J$ in the initial and final states. We decompose this
matrix element in the set of the linear independent matrices
:  \begin{equation}
D^J(p_c,p_c')\left(\Gamma_\mu\left(p_c'\right)p_c^\mu\right)^n
\;,\quad n = 0,1\ldots 2J\;.
\label{Gpn}
\end{equation}
Here $\Gamma_\mu\left(p_c'\right)$ -- is the spin 4--vector
defined with the use of the Pauli--Lubanski vector (see, e.g.,
\cite{KoT72}, and also the Sec.IV(A)).

The set of matrices (\ref{Gpn}) is the set of Lorentz scalars
(scalars and pseudoscalars). The decomposition contains
the 4--vectors analogous to
$B_\mu$ ¢ (\ref{j=BG}).

We used the described approach to consider the systems with
$J$ = 1 ($\rho$ --meson, deuteron) and obtained good description
of the experimental data
\cite{KrT01}.

Now let us proceed with the approximate calculation of the form factor
(\ref{int G=Fá}).

\subsection*{C.
Modified impulse approximation
(MIA)}

The problem of the calculation of the form factor
$G(s\,,Q^2\,,s')$ (\ref{int G=Fá}) including exchange currents
is a very difficult problem. We propose an approximation which
is a kind of analog of relativistic impulse approximation. We
propose to omit the contribution of the two--particle currents
to the form factor $G(s\,,Q^2\,,s')$.

However we will not change the covariant part
$B_{\mu}$ of the current matrix element in
(\ref{j=BG}), so that this covariant part will contain the
contribution of the two--particle currents and so that the
transformation properties of the matrix element will not be
changed.

So, we approximately change the generalized function
$G(s\,,Q^2\,,s')$
in (\ref{j=BG}), (\ref{G=g0+Gc})
for the generalized function $g_0 (s,Q^2,s')$
(\ref{j=A mu g0}), (\ref{ff-nonint}),
which describes, as we have shown before, the electromagnetic
properties of the free two--particle system. Nevertheless,
the matrix element
(\ref{int=Fá}), (\ref{j=BG})
as a whole will contain the contributions of two--particle
currents, although not the full contribution but such that
ensures its correct transformation properties.

Let us note that our approximation does not contradict general
statements (see ~\cite{KeP91}) that to obtain correct
description of electromagnetic current of composite system which
satisfy the Lorentz--covariance condition and the current
conservation law one has to take into account many--particle
currents.

Thus, in our approximation the scalar equality
(\ref{int G=Fá}) transforms into approximate scalar equality
which corresponds, from the physical point of view, to
relativistic impulse approximation. In the developed
mathematical formalism we have not broke the Lorentz covariance
of the current nor the current conservation law.
Let us point out that to calculate form factor
we do not use a special current component as it is
done in other mathematical formulations of RHD
(see, e.g., \cite{ChC88}). Let us remark that, from the physical
point of view, the form factor $g_0(s,Q^2,s')$ contains the
contributions of one--particle currents only
(see Equations (\ref{j=A mu g0}),
(\ref{jP=jp1p2}), (\ref{ff-nonint})) and in this sense our
approximation corresponds to the known impulse approximation.
In order to emphasize that our approximation differs from the
usual IA we will refer to it as to modified impulse
approximation (MIA). The form factor of the composite system
in MIA has the form:
\begin{equation}
F_c(Q^2) =
\int\,d\sqrt{s}\,d\sqrt{s'}\,\varphi(s)\,g_0(s\,,Q^2\,,s')
\varphi(s')\;.
\label{Fc fin}
\end{equation}

We do not discuss in this paper the problem of going beyond the
limits of MIA and of obtaining corrections to
$g_0(s\,,Q^2\,,s')$ in (\ref{G=g0+Gc}), (\ref{Fc fin}).
This means that if considering, for example, nucleon systems we
do not take into account meson current.

Let us consider now the fulfilling of the conditions (i)--(vi)
for the electromagnetic current.

The conditions (i)--(iii) are satisfied by construction. For
example the fulfilling of (i) and (iii) is ensured by the
correct transformation properties of the 4-vectors in
(\ref{j=A mu g0}), (\ref{j=BG}), and (\ref{B}).

The condition (iv) is satisfied immediately as
the form factor $g_0(s\,,Q^2\,,s')$ in (\ref{j=A mu g0})
and the form factor $G(s\,,Q^2\,,s')$ in (\ref{j=BG})
are scalars in our simple model
\footnote{
The currents which do not conserve the parity also can be
considered in our formalism. In that case one can separate not
only the scalar part of current matrix element but the
pseudoscalar part, too. This case is considered elsewhere.
}.

The condition of cluster separability (v) needs a more detailed
consideration. At large distances (or if the interaction is
switched off) the contribution of two--particle currents has to
go to zero:
$G_c(s\,,Q^2\,,s') \to 0$  in (\ref{G=g0+Gc}).
This means that in the form (\ref{G=g0+Gc})
the form factor $G(s\,,Q^2\,,s')$ has to transform into
$g_0(s\,,Q^2\,,s')$.  Let us remark that the condition of
cluster separability is fulfilled in MIA, too, as in this
approximation the use of $g_0(s\,,Q^2\,,s')$ instead of
$G(s\,,Q^2\,,s')$ is supposed from the very beginning. When the
interaction is switched off the generalized function
$g_0(s\,,Q^2\,,s')$ for the free two--particle system acts on
a larger space of test functions than
(\ref{psi(ss')}). As $g_0(s\,,Q^2\,,s')$ contains only
the one--particle current contributions
(\ref{jP=jp1p2}) the condition (v)
is satisfied and the composite--system current go to the sum of
the one--particle currents.

The condition on the charge to be nonrenormalizable also is
fulfilled directly in MIA because the weak limit
(\ref{Q2=0}) does exist on test functions (\ref{psi(ss')}).

So, our prescription for the construction of the current in MIA
satisfies all the conditions for the current operator.

Let us note that the equation
(\ref{Fc fin}) for the composite-system form factor is analogous
to the equations obtained in the framework of the dispersion
approach
\cite{TrS69,ShT69}  (see also~\cite{AnK92})
based on the analytic properties of the scattering amplitudes,
matrix elements, form factors in the complex energy plane.

As the dispersion approach is
rather correctly derived in the frame of QFT
\cite{Bar65}, this fact can be considered as a possible link
between QFT and RHD. The establishment of such a link is one of
the unsolved problems of RHD
\cite{KeP91}.

Let us note that an immediate application of the
approach to quark systems is difficult to realize because of the
fact of quark confinement.  However, there are some
investigations based on similar ideas where the form factors of
hadrons as constituent--quark bound states are considered in the
frame of the dispersion technique of the integral over composite
particle mass ~\cite{Mel94}.

\subsection*{D. MIA {\it versus} IA}

Let us compare the approximation MIA with the well known IA.

To do this let us first calculate the form factor in IF RHD
not using the canonical parameterization. In particular, let us
formulate IA in terms of operators as it is formulated usually
(not in terms of form factors). Let us decompose the matrix
element (\ref{j=Fc}) through the complete set of states
(\ref{p1m1p2m2}):
$$
\langle\,p_c\,|j_\mu(0)|\,p_c'\,\rangle  =
\int\,\frac{d\vec p_1\,d\vec p_2} {2p_{10}\,2p_{20}}
\frac{d\vec p_1\,'\,d\vec p_2\,'}{2p_{10}'\,2p_{20}'}\,
$$
$$
\times \langle\,p_c\,|\vec p_1;\vec p_2\,\rangle
\langle\vec p_1;\vec p_2\,|j_\mu|\,\vec p_1\,';\vec p_2\,'\rangle
$$
\begin{equation}
\times \langle\,\vec p_1\,';\vec p_2\,'|p_c'\rangle\;.
\label{jc=j1+j2}
\end{equation}
Here $\langle\,\vec p_1;\vec p_2|p_c\rangle$ is wave function of constituents
in composite system.

If the current matrix element in ~(\ref{jc=j1+j2})
is taken in the IA approximation
~(\ref{j=jk+jkm}) and contains one--particle currents only, then
the Eq. ~(\ref{jc=j1+j2}) is selfcontradicting
~\cite{KeP91}.

To write the form factor in terms of wave functions
(\ref{wf}) one has to perform the CG decomposition of the basis
(\ref{p1m1p2m2}) in terms of the basis (\ref{PkJlSm})
in the wave functions (\ref{jc=j1+j2}) and to use the explicit
form for CG coefficients  (\ref{Klebsh}) for the quantum numbers
of the system:
\begin{equation}
\langle\,\vec p_1;\vec p_2\,|p_c\rangle  = \sqrt{\frac{2}{\pi}}\,
\langle\,\vec P\,,\sqrt{s}\,,J\,,l\,,S\,,m_J\,|p_c\rangle .
\label{wfp1p2}
\end{equation}

The current matrix element in
(\ref{jc=j1+j2}) has the form (\ref{j=j1xI}).
The one--particle currents are expressed through the form
factors (\ref{j=f1}).

The Eq.(\ref{jc=j1+j2}) is an equality for two 4--vectors.
Taking different components of this equality and exploiting
$\delta$--functions in integrals, one can calculate the form
factor of the composite system. The result of calculation of the
form factor in this way is not unambiguous. In particular, it
depends on the actual choice of the component of the current
(\ref{jc=j1+j2})
to be used in the calculation.
Moreover, the result depends on the coordinate frame chosen to
perform the integration in
(\ref{jc=j1+j2}). This is the general feature of IA in the usual
formulation of IF RHD (see, e.g., \cite{KeP91}).

Let us write the final result of the calculation of the form
factor from the equation for the null--component of the current
and performing the integration in the coordinate frame where
$\vec p_c\,' = 0\>,\>\vec p_c = (0,0,p)$.
If now we write the integral in terms of the invariant variables
$s,s'$ the obtained form factor has the form:
$$
F_c(Q^2) =
\frac{M_c}{4}\frac{\sqrt{2\,(2\,M_c^2 + Q^2)}} {4\,M_c^2 + Q^2}\,
$$
$$
\times
\int\,\sqrt{\frac{s}{s'}}\,\frac{d\sqrt{s}\,
d\sqrt{s'}}{\sqrt{(s - 4\,M^2)(s' - 4\,M^2)}}
\frac{(s + s' + Q^2)^4\,Q^2}{[\lambda(s\,,-\,Q^2\,,s')]^{3/2}}
$$
\begin{equation}
\times
\frac{1}{(s\,s')^{1/4}}\frac{\theta(s,Q^2,s')}{\sqrt{s'}\,
(s + Q^2)}\,\varphi(s)\, \varphi(s')\,f_1(Q^2)\;.
\label{Fc conv}
\end{equation}
The Eq.(\ref{Fc conv}) differs from (\ref{Fc fin}),
obtained with the use of the two--particle free form factor. In
the case of wave functions satisfying the conditions
(\ref{phi(s)}), (\ref{norm}), the form factor (\ref{Fc conv})
satisfies the normalization: $F_c(0) = e_c$.
Let us note that the form factor obtained in this way from the
third current component in (\ref{jc=j1+j2}) does not satisfy
this condition.

Let us compare IA and MIA results.
Let us note once again that in MIA we separate (by use of the
scheme of canonical parameterization) the covariant part of the
current matrix element in
(\ref{int BG=Fá}) prior to perform any calculations. This
covariant part ensures the correct transformation properties of
the corresponding decompositions in terms of free--particle
states.
The difference between
(\ref{Fc fin}) and (\ref{Fc conv}) is:
$$
\Delta F_c(Q^2) =
\int\,d\sqrt{s}\,d\sqrt{s'}\,\,\varphi(s)\,\varphi(s')\,
$$
\begin{equation}
\times g_0(s,Q^2,s')\,\left[\,1 - R(s,Q^2,s')\right]\;.
\label{DFc}
\end{equation}
$$
R(s,Q^2,s') =
\frac{M_c}{2}\frac{\sqrt{2\,(2\,M_c^2 + Q^2)}} {4\,M_c^2 + Q^2}\,
\sqrt{\frac{s}{s'}}\,
$$
\begin{equation}
\times \frac{(s + s' + Q^2)^2}{(s\,s')^{1/4}}
\frac{1}{\sqrt{s'}\,(s + Q^2)}\;.
\label{R}
\end{equation}

The value $R(s,Q^2,s')$ presents an additional factor to
one--particle currents, that is in reality the two--particle
current contributions. This term ensures the Lorentz covariance
of the electromagnetic current matrix element and the current
conservation law in  (\ref{int dmu=Fá}). Let us note that this
additional term contains no dynamical information on the
interaction of test particle with two constituents
simultaneously. It does not depend, for example, on the
interaction constants for such a process.

So, to summarize, we can write the following schematic
equations:
$$
(IA)_{Breit} \ne (IA)_{Lab}
$$
$$
(MIA)_{Breit} = (MIA)_{Lab}
$$
It is well known that the standard IA depends strongly on the
coordinate frame used for the calculation. The MIA results
do not depend on it at all. So, the differences between IA and
MIA results for different IA coordinate frame can be
rather significant.

Notice that IA and MIA coincide in the nonrelativistic limit. As
this takes place, the nonrelativistic limits of form factors,
which were obtained from the different current components, are
identical. Hence the difference between IA and MIA is connected
with the breaking of relativistic covariance conditions really.

We give the quantitative comparison of the
form factors obtained in IA and in MIA in the Section IV
where the realistic calculation of the pion electromagnetic
structure is given.

\subsection*{E. The nonrelativistic limit}

The description of composite--system form factors in terms of
distributions is not a specific feature of our relativistic
approach. The similar formalism is used in nonrelativistic
theory of composite systems \cite{BrJ76} for a rather long
time (although not referring to the mathematics of
distributions). In the nonrelativistic limit our approach gives
the formalism developed in \cite{BrJ76}.

In the nonrelativistic limit the relativistic charge form factor
(\ref{Fc fin}) has the following form:
\begin{equation}
F_{NR}(Q^2) =
\int\,k^2\,dk\,k'\,^2\,dk'\,u(k)\,g_{0NR}(k,Q^2,k')\,u(k')\;,
\label{FNR}
\end{equation}
\begin{equation}
g_{0NR}(k,Q^2,k') = \frac{f_1(Q^2)}{k\,k'\,Q}\theta(k,Q^2,k')\;,
\label{g0NR}
\end{equation}
$$
\theta(k,Q^2,k') = \vartheta\left(k' - \left|k - \frac{Q}{2}\right|\right) -
\vartheta\left(k' - k - \frac{Q}{2}\right)\;.
$$
Here $g_{0NR}(k,Q^2,k')$ is the free relativistic form factor
obtained from (\ref{ff-nonint}) in the nonrelativistic limit.
$f_1(Q^2)$ is the charged--particle form factor.
The obtained result coincides with that derived in standard
nonrelativistic calculations \cite{BrJ76}.

Rigorously speaking, the Eq.(\ref{FNR})
has to be interpreted as a functional in the sense of
distributions generated by the function
$g_{0NR}(k,Q^2,k')$ and defined on  test functions
$u(k)\,u(k')$.  The ordinary function
(\ref{g0NR}) generates regular
generalized function defined generally on the larger
class of test functions
$\psi(k,k')$ in {\bf R}$^2$, providing the uniform convergence of the
integral. One needs the uniform convergence to take limits
in the integrands.

Let us define the functional in {\bf R}$^2$
by the following regular distribution (compare with
(\ref{<>})--(\ref{dmu})):
$$
\langle\,g_{0NR}(k,Q^2,k')\,,\,\psi(k,k')\rangle
$$
\begin{equation}
= \int\,d\mu (k,k')\,g_{0NR}(k,Q^2,k')\,\psi(k,k')\;,
\label{<g0NR>}
\end{equation}
$$
d\mu (k,k') = \theta(k)\,\theta(k')d\mu (k)\,d\mu (k')\;,
\quad d\mu (k) = k^2\,dk\;.
$$

The function $g_{0NR}(k,Q^2,k')$ which appears in
\cite{BrJ76} quite formally, here has a definite physical
meaning and describes the electromagnetic properties of
nonrelativistic free system of two spinless particles in the
$S$ -- state, one of particle having no charge (compare with
$g_0(s,Q^2,s')$ in (\ref{j=A mu g0}), (\ref{ff-nonint}),
(\ref{<>})).  The statical limit
$\lim_{Q^2\to 0}\,g_{0NR}(k,Q^2,k')$
giving the system charge exists only in the weak sense as the
limit of the functional (\ref{<g0NR>}):
$$
\lim_{Q^2\to 0}\,\langle\,g_{0NR}(k,Q^2,k')\,,\,\psi(k,k')\rangle
$$
\begin{equation}
= \langle\,e_c\,\delta(\mu(k') - \mu(k)),\psi(k,k')\rangle\;.
\label{slimNR}
\end{equation}

On the test functions $\psi(k,k') = u(k)\,u(k')$ (with
$u(k)$ -- being the normalized bound state wave function),
the functional (\ref{<g0NR>}) defines the bound state form
factor in the nonrelativistic IA
(\ref{FNR}).  The weak limit (\ref{slimNR})
is equal to the system charge:
$$
\lim_{Q^2\to 0}\,\langle\,g_{0NR}(k,Q^2,k')\,,\,\psi(k,k')\rangle
$$
\begin{equation}
= e_c\,\int_0^\infty\,k^2\,dk\,u^2(k) = e_c\;.
\end{equation}

To go beyond nonrelativistic IA one has to addend some terms
to $g_{0NR}(k,Q^2,k')$. For example, such terms cause the meson
exchange currents in two--nucleon systems. So, in the standard
nonrelativistic theory the dynamical treatment of exchange
currents is performed in the same way as in our relativistic
approach (\ref{G=g0+Gc}).

So, to conclude, one can consider our approach to IA to be a
relativistic generalization of nonrelativistic IA, and our
equations for form factors in this approximation to be a
relativistic generalization of the equations of
\cite{BrJ76}. Let us remark that in more complicated systems
(e.g., in $\rho$ -- meson and deuteron) our relativistic form factors
also have correct nonrelativistic limits which coincide with
\cite{BrJ76}.


\section{The electroweak structure of pion}

Now we apply the method of previous sections to the calculation
of the electroweak structure of pion. There exists a lot of
experimental data on pion, so the effectiveness of the method can
be checked by the comparison with the data (see, e.g., \cite{Jau91} and
references therein).

\subsection*{A. The electromagnetic form factor of pion}

The pion is spinless, so the electromagnetic current matrix
element has the form
(\ref{j=Fc}) with  $p_c\to p_\pi\;,\;F_c(Q^2)\to F_\pi(Q^2)$. In the frame of
composite quark model pion is considered as the bound state of
$u$-- and $\bar d$-- quarks. We suppose that quark masses are
equal: $m_u = m_d = M$.

To calculate in MIA the composite--system form factor one needs
to construct first the free two--particle form factor
(\ref{j=A mu g0}), (\ref{ff-nonint}), (\ref{Fc fin}).
Contrary to the simple model of the previous Section now we
consider the system of two charged particles with spins 1/2.
This gives the following complications. First, the Eq.
(\ref{j=j1xI}) for the current operator of the free system is
now transformed to the form:
\begin{equation}
j^{(0)}_\mu(0) =
j_{1\mu}\otimes I_{2}\oplus j_{2\mu}\otimes I_{1}\;.
\label{j=j1xI+j2xI}
\end{equation}
Here $j_{(1,2)\mu}$ - the electromagnetic currents of
particles,
$I_{(1,2)}$ -- the unity operators in the one--particle state
Hilbert spaces.
The Eq.(\ref{j=j1xI+j2xI}) can be rewritten in terms of matrix
elements:
$$
\langle\vec p_1,m_1;\vec p_2,m_2|j_\mu^{(0)}(0)|
\vec p\,'_1,m'_1;\vec p\,'_2,m'_2\rangle  =
$$
\begin{equation}
= \langle\vec p_2,m_2|\vec p\,'_2,m'_2\rangle
\langle\vec p_1,m_1|j_{1\mu}|\vec p\,'_1,m'_1\rangle
+ (1\leftrightarrow 2)\;.
\label{j=j1+j2}
\end{equation}

Second, the matrix element of one--particle current contains
now, contrary to (\ref{j=f1}), the magnetic form factors of
quarks as well as the charge ones.
Now the parameterization (the elementary--particle one following
\cite{KeP91}) is of the form:
$$
\langle\vec p,\,m|j^\mu(0)|\vec p\,',\,m'\rangle
$$
\begin{equation}
= \overline u_{\vec p\,m}\gamma ^\mu u_{\vec p\,'m'}\,F_1(Q^2) -
\overline u_{\vec p\,m}\sigma ^{\mu \nu}q_\nu \,u_{\vec p\,'m'}\,F_2(Q^2)\;,
\label{j=F1F2}
\end{equation}
$u_{\vec pm}$ - the Dirac bispinor,
$\gamma^\mu$ - Dirac matrix,
$$
\sigma^{\mu \nu} =
\frac{1}{2}(\gamma^\mu\gamma^\nu - \gamma^\nu\gamma^\mu)\;,\quad
q_\nu = (p - p')_\nu\;,
$$
Using multipole parameterization we can write the one--particle
current matrix element in terms of Sachs form factors:
$$
G_E(Q^2) =
\tilde F_1(Q^2) + \frac {\kappa Q^2}{4M^2}\,\tilde F_2(Q^2)\;,
$$
$$
G_M(Q^2) = \tilde F_1(Q^2) + \kappa \tilde F_2(Q^2)\;,
$$
\begin{equation}
F_1(Q^2) = e\tilde F_1(Q^2)\;,\quad
F_2(t) = \frac {\kappa }{2M} \tilde F_2(Q^2)\;.
\label{G=F}
\end{equation}
Here $G_{E,M}$ - Sachs electric and magnetic form factors,
respectively,
$e$ is the particle charge,
$\kappa $ is the anomalous magnetic moment.

It is convenient to use the canonical parameterization of matrix
elements \cite{ChS63}:
$$
\langle\,\vec p,\,m\,|\,j_\mu(0)\,|\,\vec p\,',\,m'\,\rangle
$$
$$
= \sum_{m''}\,\langle m|D^j(p,\,p')|m''\rangle
\langle m''|\,f_1(Q^2)K'_\mu + i\,f_2(Q^2)R_\mu|m'\rangle\;,
$$
\begin{equation}
K'_\mu = (p + p')_\mu \,, \quad
R_\mu = \epsilon _{\mu \,\nu \,\lambda \,\rho}\, p^\nu
\,p'\,^\lambda \,\Gamma^\rho (p')\;.
\label{j=f1f2}
\end{equation}
$\Gamma(p)$ is 4--vector of spin:
$$
\vec \Gamma(p) =  M\,\vec j + \frac {\vec p(\vec p\vec j)}{p_0 + M}\;,\quad
\Gamma_0(p) = (\vec p\vec j)\;.
$$
The form factors
$f_1(Q^2)$, $f_2(Q^2)$ are the electric and magnetic form
factors of particles. They are connected with Sachs form
factors \cite{BaY95}:
$$
f_1(Q^2) = \frac {2M}{\sqrt {4M^2 + Q^2}}\,G_E(Q^2)\;,
$$
\begin{equation}
f_2(Q^2) = -\frac {4}{M\sqrt {4M^2 + Q^2}}\,G_M(Q^2)\;.
\label{Bal}
\end{equation}

Third, now the CG coefficients are of more complicated form.
They are given by
(\ref{Klebsh}) with $J = S = l = 0$. Contrary to the previous
simple case, now the CG coefficients contain the Wigner rotation
matrices.

Finally, the free two--particle form factor for the system of
two particles with spin 1/2 and quantum numbers
$J = S = l = 0$ is of the form (see also \cite{BaK96}):
$$
g^{q\bar{q}}_0(s,Q^2,s')=
n_c\,\frac{(s+s'+Q^2)Q^2}{2\sqrt{(s-4M^2) (s'-4M^2)}}\;
$$
$$
\times \frac{\theta(s,Q^2,s')}{{[\lambda(s,-Q^2,s')]}^{3/2}}
\frac{1}{\sqrt{1+Q^2/4M^2}}
$$
$$
\times \left\{(s+s'+Q^2)(G^u_E(Q^2)+G^{\bar d} _E(Q^2))\cos(\omega_1+\omega_2)
+\right.
$$
\begin{equation}
+ \left.\frac{1}{M}\xi(s,Q^2,s')
(G^u_M(Q^2)+G^{\bar d}_M(Q^2)) \sin(\omega_1 + \omega_2)\right\}\;,
\label{g_0}
\end{equation}
Here
$$
\xi(s,Q^2,s')=\sqrt{ss'Q^2-M^2\lambda(s,-Q^2,s')}\;,
$$
$n_c$ is the number of quark colours,
$\omega_1$ ¨ $\omega_2$ -- the Wigner rotation parameters:
$$
\omega_1 =
\arctan\frac{\xi(s,Q^2,s')}{M\left[(\sqrt{s}+\sqrt{s'})^2
+ Q^2\right] + \sqrt{ss'}(\sqrt{s} +\sqrt{s'})}\;,
$$
\begin{equation}
\omega_2 = \arctan\frac{
\alpha (s,s') \xi(s,Q^2,s')} {M(s + s' + Q^2)
\alpha (s,s') + \sqrt{ss'}(4M^2 + Q^2)}\;,
\label{omega}
\end{equation}
with $\alpha (s,s') = 2M + \sqrt{s} + \sqrt{s'}$, and
$G^{u,\bar d}_{E,M}(Q^2)$ are Sachs form factors for quarks.
The $\theta$ -- function in (\ref{g_0}) is the same as in (\ref{ff-nonint}).

An interesting effect follows from
(\ref{g_0}): due to the relativistic Wigner spin rotation
effect the pion charge form factor contains the contribution of
quark magnetic form factors.

The pion charge form factor can be calculated using
(\ref{Fc fin}), with
(\ref{g_0}) for the free two--particle form factor:
\begin{equation}
F_\pi(Q^2) = \int\,d\sqrt{s}\,d\sqrt{s'}\,
\varphi(s)\,g^{q\bar{q}}_0(s\,,Q^2\,,s') \varphi(s') .
\label{Fpi}
\end{equation}

\subsection*{B. The lepton decay constant of pion}

The lepton decay constant $f_\pi$ is defined by the
electroweak--current matrix element
\cite{Jau91}:
\begin{equation}
\langle0|j_\mu(0)|\,p_\pi\,\rangle  =
if_\pi\,p_\pi\,_\mu\frac{1}{(2\pi)^{3/2}}\;.
\label{j=f_c}
\end{equation}
$p_\pi$ -- 4-momentum of meson. Let us decompose the l.h.s.
of (\ref{j=f_c}) in the basis (\ref{PkJlSm}). Using the explicit
form of the meson wave function
(\ref{wf}) one can obtain for (\ref{j=f_c}):
$$
\int\,\frac{N_c}{N_{CG}}\,d\sqrt{s}\,
\langle 0|j_\mu(0)|\vec p_\pi\,,\sqrt{s}\rangle \varphi(s)
$$
\begin{equation}
= if_\pi\,p_\pi\,_\mu\frac{1}{(2\pi)^{3/2}}\;.
\label{int ds=f_c}
\end{equation}
As in Section II (Eq.(\ref{j=BG})) one can divide the integrand
in (\ref{int ds=f_c}) into two parts:
the covariant part (smooth ordinary function) and the invariant
part.
\begin{equation}
\frac{N_c}{N_{CG}}\, \langle0|j_\mu(0)|\vec p_\pi\,,\sqrt{s}\rangle
= iG(s)B_\mu(s)\frac{1}{(2\pi)^{3/2}}\;.
\label{j=G(s)}
\end{equation}

The invariant form factor $G(s)$ is a generalized function. In
the same way as in calculating
(\ref{int G=Fá}) of the previous section, we now obtain the
lepton decay constant of pion in the form
\begin{equation}
\int\,d\sqrt{s}\,G(s)\varphi(s) = f_\pi\;.
\label{int G(s)=f_c}
\end{equation}

In general, the form factor
$G(s)$ can be calculated in the frame of the standard model for
electroweak interactions. However, in this paper we limit
ourselves by 4-fermion interaction. We take for
$G(s)$  the form factor which parameterizes the decay of free
two--quark system:
\begin{equation}
\langle0|j^{(0)}_\mu(0)|\vec P\,,\sqrt{s}\rangle  =
iG_0(s)P_\mu\frac{1}{(2\pi)^{3/2}}\;.
\label{j0=G0}
\end{equation}
The explicit form
(\ref{j0=G0}) is written by analogy to
(\ref{j=A mu g0})   not taking into account the current
conservation law. The form
(\ref{j0=G0}) is quite similar to
(\ref{j=f_c}) but instead of the constant
$f_\pi$ the form factor depending on invariant variables
is written.
To calculate $G_0(s)$ let us decompose (\ref{j0=G0}) in the
one--particle basis (\ref{p1m1p2m2}).  Now we obtain for
(\ref{j0=G0}):
$$
iG_0(s)P_\mu\frac{1}{(2\pi)^{3/2}}
$$
$$
= \sum_{m_1\,,m_2\,,i_c}\int\,\frac{d\vec p_1}{2p_{10}}\,
\frac{d\vec p_2}{2p_{20}}\,
\langle 0|j^{(0)}_{\mu\,i_c}|\vec p_1\,,m_1\,;\vec p_2\,,m_2\rangle
$$
\begin{equation}
\times \langle\vec p_1\,,m_1\,;\vec p_2\,,m_2|\vec P\,,\sqrt{s}\rangle .
\label{G0=int}
\end{equation}
$i_c =1,2,3$, the sum over $i_c$ is the sum over the colours.
The CG coefficients are known (\ref{Klebsh}).
The current matrix element in the basis
(\ref{p1m1p2m2}) can be written in the standard way in terms of
the lepton decay current matrix element
\cite{Jau91}:
$$
\langle 0|j^{(0)}_\mu|\vec p_1\,,m_1\,;\vec p_2\,,m_2\rangle
$$
\begin{equation}
=
\frac{1}{(2\pi)^{3}}\,\bar v(\vec p_2\,,m_2)
\gamma_\mu(1 + \gamma^5) u(\vec p_1\,,m_1)\;.
\label{j0=vgamu}
\end{equation}
We integrate in (\ref{G0=int}) in the coordinate frame with
$\vec P =0$.  Finally, we obtain:
\begin{equation}
G_0(s) =
\frac{n_c}{2\sqrt{2}\,\pi\,P_0} (p_{0}+M)\,
\left[1 - \frac{k^2}{(p_{0}+M)^2}\right]\;,
\label{G0}
\end{equation}
$$
p_{0} = \sqrt{k^2 + M^2}\;.
$$

Substituting (\ref{G0})
in the Eq.(\ref{int G(s)=f_c}) we obtain the result which has
the following form if written in invariant variables:
\begin{equation}
f_\pi =
\frac{2M\,n_c}{2\sqrt{2}\,\pi}\,\int\,d\sqrt{s}
\frac{1}{\sqrt{s}}\, \varphi(s)\;.
\label{fpi}
\end{equation}

Let us notice that the Eq.(\ref{fpi}) coincides with that
obtained in the frame of light--front dynamics
\cite{Jau91}. However, although all forms of RHD are unitary
equivalent
\cite{Lev95}, nevertheless after the physical approximations
are made in more complicated cases
the results, e.g. for form factors, can be
different.  This is possibly due to the fact that the unitary
operators connecting different forms of RHD are interaction
dependent \cite{Lev95} and so the RHD forms realize one and the
same approximation in different ways.

Let us remark that the nonrelativistic limit of the
Eq.(\ref{fpi}) gives the standard form in terms of coordinate
space wave function at zero value.


\subsection*{C. The results of calculations}

To calculate the electroweak structure of pion using
(\ref{Fpi}), (\ref{g_0}), (\ref{fpi}), (\ref{phi(s)})
the following meson wave functions were utilized:

1. A gaussian or harmonic oscillator (HO)  wave function
\begin{equation}
u(k) = N_{HO}\,\hbox{exp}\left(-{k^2}/{2b^2}\right).
\label{HO-wf}
\end{equation}

2. A power-law (PL)  wave function
\begin{equation}
u(k) = N_{PL}\,{(k^2/b^2 + 1)^{-n}}\>,\quad n = 2\>,3\;.
\label{PL-wf}
\end{equation}

3. The wave function with linear confinement from
Ref.\cite{Tez91}:
$$
u(r) = N_T \,\exp(-\alpha r^{3/2} - \beta r)\>,\quad
\alpha =\frac{2}{3}\sqrt{M\,a}\>,\quad
$$
\begin{equation}
\beta = \frac{M}{2}\,b\;.
\label{Tez91-wf}
\end{equation}
$a\>,b$ -- parameters of linear and Coulomb parts of potential
respectively.

In the Ref.\cite{BaK96} in the calculation of pion
electromagnetic structure we supposed the quarks to be
point--like. The results of \cite{BaK96} can be considered as
preliminary results. However, one has to take into account the
structure of constituent quarks \cite{Ger8995}, in particular,
the anomalous magnetic moment. As anomalous magnetic moments
are connected with finite size of quark, one has to take into
account the explicit form of quark form factors entering
(\ref{g_0}) and the pion charge form factor
(\ref{Fpi}).
As in \cite{CaG96} let us use the following forms for quark form
factors:
$$
G^{q}_{E}(Q^2) = e_q\,f(Q^2)\;,
$$
\begin{equation}
G^{q}_{M}(Q^2) = (e_q + \kappa_q)\,f(Q^2)\;.
\label{q ff}
\end{equation}
Here $e_q$ -- the quark charge, $\kappa_q$ -- the quark
anomalous magnetic moment (in natural units). To obtain the
explicit form of the function
$f(Q^2)$ let us consider the asymptotics of pion charge form
factor as
$Q^2\;\to\;\infty\;,\;M\;\to\;$0.

To obtain the asymptotic behavior let us first make the
asymptotic estimation of the integrals in (\ref{Fpi}) in the
point--like quark approximation
($f(Q^2) = 1\;,\;\kappa = 0$ in(\ref{q ff}) ).
Omitting the details of calculation (given in
\cite{KrT98}) we write the final result for the asymptotics in
the form:
\begin{equation}
F_\pi(Q^2)\quad\sim\quad Q^{-2}\;.
\label{as-qcd}
\end{equation}
The asymptotics does not depend on the actual form of the wave
function and coincides with that obtained in QCD. The actual
form we obtain, e.g. for
(\ref{HO-wf}) is:
\begin{equation}
F_\pi(Q^2)\>\sim 32\sqrt{2}
\frac{\left[\Gamma\left(\frac{5}{4}\right)\right]^2}{\sqrt{\pi}}
\frac{b^2}{Q^2}\;.
\label{as s pov}
\end{equation}
It is worth to compare the form
(\ref{as s pov}) with the detailed QCD result
\cite{Rad98}:
\begin{equation}
F_\pi(Q^2) = \frac{8\,\pi\,\alpha_s\,f_\pi^2}{Q^2}\;.
\label{Fpi qcd}
\end{equation}

If $\alpha_s/\pi\;\sim\;0.1$
then (\ref{as s pov}) and (\ref{Fpi qcd}) coincide at $b\;\sim\;$0.1.
So the asymptotics (\ref{as-qcd}) is quite realistic.

In the case of non--point--like quarks we obtain another
asymptotics because the form factor depends upon the momentum
transfer. It is known that QCD gives logarithmic corrections
to (\ref{as-qcd}). To agree with this QCD corrected
asymptotics we can, for example, choose the following form
for $f(Q^2)$:
\begin{equation}
f(Q^2) = \frac{1}{1 + \ln(1+ \langle r^2_q\rangle Q^2/6)}\;.
\label{f_qour}
\end{equation}
Here $\langle r^2_q\rangle$ is the MSR of
the constituent quark which can be considered as the model parameter. Let us
fix it (as in \cite{CaG96}) to be:
$\langle r^2_q\rangle \simeq 0.3/M^2$.

For the constituent quark mass in pion we use the value which is
usually used in the calculations in RHD: $M = $ 0.25 GeV.

The quark anomalous magnetic moments can be taken from
\cite{Ger8995}:  $\kappa_u = 0.029\;,\;\kappa_d = -\,0.059$.

We choose the parameters
$b$  in (\ref{HO-wf}), (\ref{PL-wf}) and $a$ in (\ref{Tez91-wf})
in such a way as to fit the pion MSR:
$\langle r_\pi^2\rangle  = (\hbox{0.432}\pm\hbox{0.016 ) Fm}^2\>$
\cite{Ame84}.
We choose this way to fix the model parameters because the pion
MSR is defined by the form factor at small values of
$Q^2$, that is the range where potential models work well.

The fit of the pion MSR gives the following parameters of the
wave functions: in the model (\ref{HO-wf})
$b$ = 0.2784 GeV, model (\ref{PL-wf}) at $n$ = 2 $b$ = 0.3394 GeV,
model (\ref{PL-wf}) at $n$ = 3 $b$ = 0.5150 GeV,  model (\ref{Tez91-wf})
$b = (4/3)\alpha_s$, $\alpha_s$ = 0.59  at light meson mass scale, $a$ =
0.0567 GeV$^2$.

The results of calculation are presented on Figs.1 and 2.

\begin{figure*}
\centerline{\epsfxsize=0.4\textwidth \epsfbox{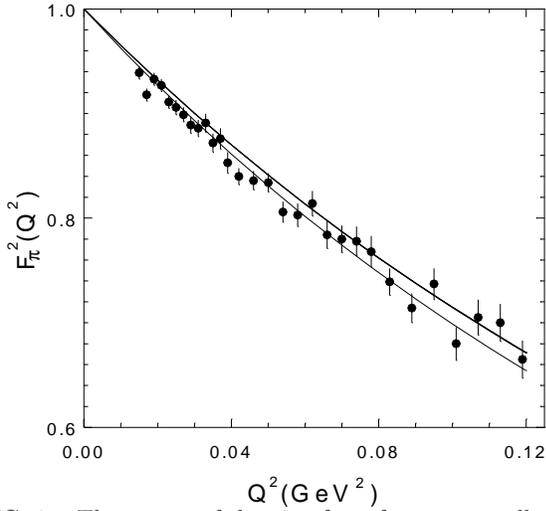}}
\caption{
The square of the pion form factor at small values of momentum
transfers for different models.}
\end{figure*}
The square of the pion form factor at small values of momentum
transfers for different models
(\ref{HO-wf}) -- (\ref{Tez91-wf}) is presented on Fig.1.
Results of
calculation in the models (\ref{HO-wf}), (\ref{PL-wf}) at $n$ = 3 and
(\ref{Tez91-wf})  coincide very closely.

The calculations of product $Q^2\,F_\pi(Q^2)$ at high
momentum transfers
for different models
(\ref{HO-wf}) -- (\ref{Tez91-wf}) are presented on Fig.2.
Legend is following: 1 -- harmonic oscillator wave function
(\ref{HO-wf}), 2 -- power--law wave function  (\ref{PL-wf}) at $n$ = 2,
3 -- power--law wave function  (\ref{PL-wf}) at $n$ = 3, and wave
function from
model with linear confinement (\ref{Tez91-wf}) (these curves coincide very
closely).

All the models for the interaction
(\ref{HO-wf}),
(\ref{PL-wf}), (\ref{Tez91-wf})
give a good description of the existing experimental data
\footnote{The JLab new results \cite{Vol00} are discussed in
connection with our approach in \cite{KrT00a}}.

The dependence of the results on the actual model is much less
pronounced that in the case of point--like quarks \cite{BaK96}.

The lepton decay constants calculated following Eq.
(\ref{fpi}) with different wave functions have the following
values:
$f_\pi =0.1210$ GeV  in the model (\ref{HO-wf}), $f_\pi =0.1327$
GeV in the model (\ref{PL-wf}) with $n=2$,
$f_\pi =0.1282$ GeV in the model (\ref{PL-wf}) with $n=3$,
and $f_\pi =0.1290$ GeV in the model (\ref{Tez91-wf}).
Let us emphasize that we have used no fitting parameters to
calculate the lepton decay constant. Nevertheless, the obtained
values are very close to the experimental value:
$f_{\pi\>exp} = 0.1317\pm 0.0002$ GeV \cite{PDG}.

\begin{figure*}
\centerline{\epsfxsize=0.4\textwidth \epsfbox{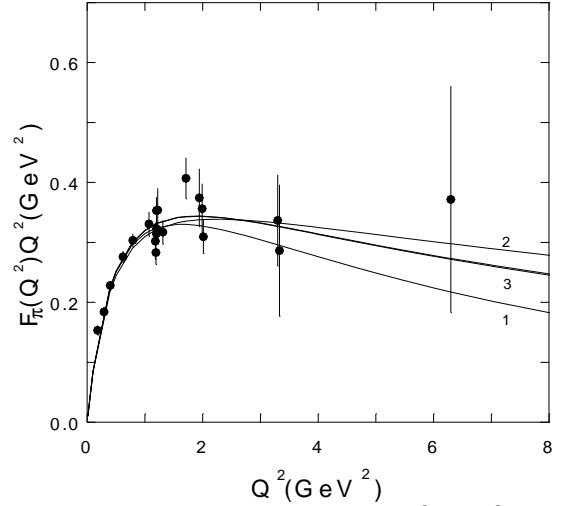}}
\caption{
Electromagnetic form factor,\protect$Q^2\,F_\pi(Q^2)$, at high
momentum transfers.}
\end{figure*}

Now let us compare the numerical results for the pion form
factor obtained in
MIA (\ref{Fpi}) with that of the traditional IA.
Let us choose for
the comparison, for example, the null--component of the current.

To obtain the pion form factor in IA we proceed in the same
way as while obtaining
(\ref{Fc conv}) of the preceding Section.
Now, however,

1) the decomposition (\ref{j=Fc})
of the IA matrix current element over the state set
(\ref{p1m1p2m2}) is realized following (\ref{j=j1+j2}),

2) the parameterization of the one--particle matrix element
is given by (\ref{j=f1f2}), (\ref{Bal}) (instead of (\ref{j=f1})),

3) the CG coefficient (\ref{Klebsh}) in (\ref{wfp1p2})
are for pion quantum numbers.

Acting in the same way as while obtaining
(\ref{Fc conv}), and using the null--component
of the current matrix element, we can write
the pion form factor in IA in the following form:
$$
F_\pi(Q^2)
= \frac{M_\pi}{4}\frac{\sqrt{2\,(2\,M_\pi^2 + Q^2)}}
{4\,M_\pi^2 + Q^2}\,\frac{n_c}{\sqrt{1 + Q^2/4M^2}}
$$
$$
\times
\int\,\sqrt{\frac{s}{s'}}\,\frac{d\sqrt{s}\,d\sqrt{s'}}
{\sqrt{(s - 4\,M^2)(s' - 4\,M^2)}}
$$
$$
\times
\frac{(s + s' + Q^2)^3\,Q^2}{[\lambda(s\,,-Q^2\,,s')]^{3/2}}
\frac{1}{(s\,s')^{1/4}}
\frac{1}{\sqrt{s'}\,(s + Q^2)}\,\varphi(s)\, \varphi(s')
$$
$$
\times
\left\{(s + s' + Q^2)\left[G^u_E(Q^2) + G^{\bar d}_E(Q^2)\right]
\cos(\omega_1 + \omega_2)\right .
$$
\begin{equation}
+
\frac{1}{M}\,\xi(s,Q^2,s')\left.
\left[G^u_M(Q^2) + G^{\bar d}_M(Q^2) \right]\sin(\omega_1 + \omega_2)\right\}.
\label{Fpi conv}
\end{equation}
Here $M_\pi$ = 139.568$\pm$0.001 MeV \cite{PDG} is mass of pion.

The normalization condition $F_\pi(0) = 1$
is satisfied for the form factor (\ref{Fpi conv})
if the wave functions (\ref{phi(s)}) satisfy
(\ref{norm nc}).

To compare the numerical results given by the
Eqs.(\ref{Fpi}), (\ref{g_0})  with that given by
(\ref{Fpi conv}) let us calculate the pion form factor using the
wave function
(\ref{HO-wf}) with the parameters of the calculations presented
in Figs.1 and 2. The results are shown in the Fig.3.
The results obtained with the use of the parameterization
(\ref{Fc fin}), (\ref{g_0}) differ essentially from that
obtained without such parameterization (\ref{Fpi conv}).
The form factor calculated in our approach describes the
existing experimental data adequately.

\begin{figure*}
\centerline{\epsfxsize=0.4\textwidth \epsfbox{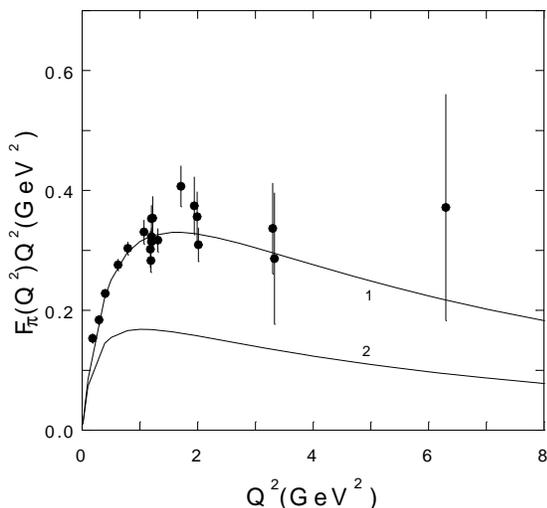}}
\caption{
$Q^2 F(Q^2)$ for MIA (1) and for IA (2). Results of calculation
with wave function (\protect\ref{HO-wf}). Parameters are the
same as in Fig.1.}
\end{figure*}

Let us emphasize once again that the form factor obtained in MIA
does not depend on the choice of coordinate frame. This is an
important advantage of our relativistic MIA.

\section{Conclusion}

Let us summarize the results.
\begin{enumerate}

\item
A new approach to the electromagnetic properties of
two--particle composite
systems is developed. The approach is based on IF RHD.

\item The main novel feature of this approach is the new method
of construction of the matrix element of the electroweak current
operator.
The electroweak current matrix element satisfies the
relativistic covariance conditions and in the case of the electromagnetic
current also the conservation law automatically.

\item The method of the construction of the current operator
matrix element consists of the extraction of the invariant part
-- the reduced matrix element on the Lorentz group (form
factor) -- and the covariant part defining the transformation
properties of the current. The form factors contain all the
dynamical information about transition. The properties of the
system as well as the approximations used are formulated in
terms of form factors, which in general have to be
considered as generalized functions.

\item The approach makes it possible to formulate relativistic
impulse approximation (modified impulse approximation --
MIA) in such  a way that the Lorentz--covariance of the current
is ensured. In the electromagnetic case the current conservation
law is ensured, too.

\item The results of the calculations are
unambiguous:  they do not depend on the choice of the coordinate
frame and on the choice of "good" components of the current as
it takes place in the standard form of light--front dynamics.

\item The effectiveness of
the approach is demonstrated by the calculation of the
electroweak structure of the pion. Our approach gives good
results for the pion electromagnetic form factor in the whole
range of momentum transfers available for experiments at present
time.

\end{enumerate}

\section*{Acknowledgements}
The authors thank V.V.Andreev and D.I.Melikhov for helpful discussions.
This work was supported in part by the Program "Russian
Universities -- Basic Researches" (grant No. 02.01.28).

\end{document}